\documentclass[10pt,a4paper]{article}

\usepackage[utf8]{inputenc}       % Input text encoding (UTF-8 for international characters)
\usepackage[T1]{fontenc}          % Font encoding for better European character support
\usepackage[english]{babel}       % Language support and localization (hyphenation, date, etc.)

\usepackage{tgheros}              % Font package providing Arial-like font
\usepackage{microtype}            % Improves general typographic appearance (kerning, spacing)

\usepackage{xcolor}               % Enhanced color support (required for \textcolor)
\usepackage{svgcolor}             % Color support specifically for SVG graphics (optional, can be omitted)
\usepackage{graphicx}             % Graphics inclusion with scaling and rotation
\usepackage{svg}                  % Including SVG vector graphics
\usepackage{float}                % Improved float placement control ([H] option)
\usepackage{wrapfig}              % Wrap text around figures and tables

\usepackage{array}                % Extended table formatting
\usepackage{booktabs}             % Professional tables with better spacing rules
\usepackage{makecell}             % Multi-line cells and custom formatting in tables

\usepackage{amsmath}              % Enhanced math typesetting and environments
\usepackage{amssymb}              % Additional math symbols from AMS
\usepackage{uniinput}             % Unicode input support and math symbols (required by your template)

\usepackage{cite}                 % Improved citation formatting and compression (collapsing ranges)

\usepackage{geometry}             % Page layout and margin control
\usepackage{tcolorbox}            % Colored boxes and frames for highlighting text
\usepackage{ragged2e}             % Provides commands for ragged text alignment

\usepackage{etoolbox}             % Programming tools and logic tests for LaTeX
\usepackage{capt-of}              % Captions in non-float environments
\usepackage{subcaption}

\usepackage{jheppubmod}           % JHEP style template (author, affiliation, title handling)

\newcommand{\bea}{\begin{eqnarray}}
\newcommand{\eea}{\end{eqnarray}}
\newcommand{\be}{\begin{equation}}
\newcommand{\ee}{\end{equation}}
\newcommand{\dd}{\mathrm{d}}
\newcommand{\e}{\mathrm{e}}

\allowdisplaybreaks

\title{Self-gravitating electromagnetic waves in the dark bubble model}

\author[A]{Ulf Danielsson}
\author[B]{, Daniel Panizo}
\author[A]{, Vincent Van Hemelryck}

\affiliation[A]{Institutionen f\"or fysik och astronomi,
Uppsala Universitet, Box 803, SE-751 08 Uppsala, Sweden}
\affiliation[B]{Department of Physics \& Astronomy, Kyoto University,
Kyoto 606-8502, Japan}

\emailAdd{ulfdanielsson@physics.uu.se}
\emailAdd{panizo@tap.scphys.kyoto-u.ac.jp}
\emailAdd{vincent.vanhemelryck@physics.uu.se}

\abstract{
We construct embeddings of gravitational and electromagnetic waves in the dark bubble scenario using pp-wave geometries in AdS$_5$, motivated by the fact that pp-wave spacetimes often provide exact solutions to the equations of motion. The setup is realised by gluing two AdS$_5$ pp-wave spacetimes across a three-brane. As an application, we analyse localised beams of light and their gravitational backreaction. Imposing suitable mixed boundary conditions in AdS$_5$, we find gravitational corrections consistent with a weakening of 4d gravity at the 5d  AdS scale.}

\preprint{UUITP-10/26}

\begin{document}

\maketitle

\newpage

\section{Introduction}
\label{sec:introduction}

The dark bubble was proposed in \cite{Banerjee:2018aa} as a way to make a positive cosmological constant natural in string theory \cite{Danielsson:2023aa, Danielsson:2023alz}. According to the model, the four-dimensional expanding cosmology rides on the boundary of a bubble mediating the decay between two  $\text{AdS}_{5}$ vacua, providing a promising framework for obtaining four-dimensional $\rho_{\Lambda}>0$ from an UV-complete construction. In \cite{Banerjee:2019aa}, it was shown that radiation can be induced by an AdS--Schwarzschild metric in the bulk, and that massive non-relativistic particles can be obtained from strings stretching along the throat of the bulk.

Neither of these options is appealing if one aims to embed the Standard Model into the dark bubble. If the stretched strings are fundamental strings, their effective mass is of order the Planck scale, and they are candidates for Planck mass black holes. The radiation induced from the bulk has an equation of state like radiation, but it is unclear if any electromagnetic degrees of freedom can be found. These results demonstrate that matter-like excitations can arise in such setups, but their relation to realistic four-dimensional interpretations remains obscure. Therefore, from a string phenomenology perspective, it would be much more natural for the degrees of freedom representing the Standard Model to be localised on the brane, with the bulk just encoding the relevant gravitational dynamics. The gravitational picture was explored in \cite{Danielsson:2022aa}, where it was shown that gravitational waves on the dark bubble uplift to higher-dimensional gravitational waves in the bulk, suggesting that gravitational dynamics are consistently captured in this braneworld construction.

An obstacle to such a construction is the way in which the energy-momentum tensor of matter on the brane enters into the effective energy-momentum tensor for four-dimensional Einstein gravity: its sign is \textit{negative} \cite{Banerjee:2020aa, Banerjee:2022aa}. This is already clear from the way that the positive cosmological constant $\Lambda_{4}$ appears. In order for the bubble to be able to nucleate, the tension of the brane needs to be less than the critical tension; that is, decreasing the tension makes the cosmological constant increase. In other words, all energy that is \textit{added} to the brane makes the effective energy density seen by four-dimensional gravity \textit{decrease}. This was also emphasised recently by ref.~\cite{Basile:2025aa} in the context of Standard Model Physics.

The solution to this problem lies in the gravitational back-reaction from the bulk. As explored in \cite{Banerjee:2020aa}, any matter on the brane is accompanied by a gravitational back-reaction in the bulk that overcompensates and makes the net result positive. This mechanism was clarified in \cite{Danielsson:2025aa}, where it was shown that a point mass sitting on the brane leads to the expected positive effective mass in the four-dimensional theory on the bubble. Additionally, it was also shown how corrections to the gravitational force appear at scales smaller than the bulk AdS scale $k$. Together, these results open the possibility of further exploring localised Standard Model matter on the brane consistently coupled to gravity.

A relevant feature of this braneworld construction is the existence of a natural hierarchy of scales. Amongst several attempts of embedding the dark bubble scenario in string theory \cite{Danielsson:2023aa,Danielsson:2023alz,Basile:2025lek}, in the model of ref.~\cite{Danielsson:2023aa}, the bulk AdS scale becomes $L \sim N^{1/2}\,\ell_{4}$, with $N$ denoting the number of D3-branes sourcing the dark bubble stringy embedding geometry. This particular scaling ensures a novel hierarchy\footnote{Also present in other dark bubble UV-realisations as in \cite{Basile:2025lek}.} where $\ell_{5} \ll \ell_{4}$, implying that gravity is substantially stronger in the four-dimensional cosmology compared to that in the bulk. This hierarchy, which is opposite to that of conventional compactifications \cite{DeWolfe_2003, Satheeshkumar_2011} and RS-like braneworld constructions \cite{Randall:1999aa, Randall:1999ab, Karch:2001aa, Antoniadis_2011}, underlies many of the distinctive features of the dark bubble model.

A first step toward realising matter living on the brane was taken in \cite{Basile:2023aa}, where electromagnetism was studied. There, the gravitational junction conditions were solved for electromagnetic waves on an expanding, spherical brane with induced four-dimensional cosmology, where the coupling of the waves to a bulk two-form field $B$ was also accounted for.\footnote{A very similar approach was also taken in \cite{Basile:2025lek}, where the Ramond-Ramond $C_{2}$ field of type IIB string theory was considered to be sourced instead of $B$.} The way in which the electromagnetic field sources a $B$-field in the bulk was also taken into account. In addition, the full solution for the metric and electromagnetic fields on the brane, as well as the metric and $B$-field in the bulk, were obtained, although the bulk backreaction was treated in an isotropic approximation. Nevertheless, this provided substantial insights: in particular, it was shown that the contribution from the $B$-field is sub-leading, while the dominant effect arises from the electromagnetic field localised on the brane.  In light of the later results in \cite{Danielsson:2025aa}, this can be understood as a direct consequence of the hierarchy $\ell_{5} \ll \ell_{4}$ discussed above.

In this paper, we will make a much more careful study of the electromagnetic field. In particular, we will consider the full time-dependent solutions and carefully match the four- and five-dimensional solutions to each other. This new approach relies fundamentally on the study of \textit{plane-fronted waves with parallel propagation}, better known as pp-waves, which describe any type of massless, propagating waves (electromagnetic, gravitational, Weyl fermions, etc.) \cite{Griffiths:1991zp, Hubeny_2002, Blau:2011, Bondi:1958aj, Roche:2022bcz}. These types of waves are known to be exact solutions to the equations of motion in arbitrary spacetime dimension. Note that any wave becomes a pp-wave in the Penrose limit, where one picks an observer following a light-like geodesic and shrinks the longitudinal directions while expanding the transverse ones \cite{Blau:2011}.

The paper is organised as follows: In section~\ref{sec:pp_wave_uplift}, we will study how four-dimensional pp-wave spacetimes can be embedded in five-dimensional analogues with AdS asymptotics.  Section~\ref{sec:phenomenology} will be devoted to comparing all previous insights to the results obtained for a point mass in \cite{Danielsson:2025aa}. In order to increase our comprehension of this novel induced strong gravity approach, we will consider a specific case with a beam of electromagnetic radiation. This will show how the gravitational response is modified at short distances such that the effective energy-momentum tensor is broadened for narrow beams. We will finally summarise our key findings, as well as elaborate on insights for future tasks to accommodate the Standard Model onto the dark bubble model, in section~\ref{sec:conclusion}.

\section{pp-waves on the dark bubble}
\label{sec:pp_wave_uplift}

\subsection{The physics of pp-waves}
\label{subsec:review_pp_waves}
Plane-fronted waves with parallel rays (pp-waves) form an important class of exact solutions to Einstein’s equations, describing gravitational waves propagating in a fixed null direction. They are characterised by the existence of a covariantly constant null vector and play a central role in both classical gravity and string theory due to their exact solvability, see e.g. refs.~\cite{Stephani:2003tm,Griffiths:2009dfa}. The four-dimensional pp-wave spacetime metric can be written as follows:
\begin{equation}
\label{eq:4d_wave_metric}
    \dd s^2 = - 2 \,\dd u \dd v -w(u,x,y)\, \dd u^2 + \dd x^2 +\dd y^2
\end{equation}
which is the so-called Brinkmann parametrisation. The coordinates $u$ and $v$ parametrise null directions and the wave is chosen to propagate in the $u$-direction.  The transverse coordinates can also be written in terms of complex coordinates $\zeta = x + i y$.
We are interested in solutions of this type to a theory of Einstein gravity, with or without a coupling to a U(1) gauge field, described by the following action:
\begin{equation}
    S = S_\mathrm{G} + S_{F}\,,
\end{equation}
where $S_\mathrm{G}$ is the standard Einstein-Hilbert action
\begin{equation}
    S_\mathrm{G}=\frac{1}{16\pi G_4} \int\dd^{4}\xi\, \sqrt{-g_4}\, R_4 \,,
\end{equation}
where $\sqrt{-g_{4}}$ stands for the determinant of the metric. The term $S_F$ represents the action of the gauge field with field strength $F$ and takes the standard Maxwell form:
\begin{equation}
        S_\mathrm{F}=-\frac{1}{16\pi G_4} \int\dd^{4}\xi \,\sqrt{-g_4}\,|F|^2\,,
\end{equation}
In both cases, with the metric \eqref{eq:4d_wave_metric}, the gauge field equations of motion can be solved for a field strength taking the plane wave form
\begin{equation}
    F = Q_1(u) \dd u \wedge \dd x + Q_2(u) \dd u \wedge \dd y\,.
\end{equation}
The stress tensors are the same with only one non-vanishing component:
\begin{equation}
    T_{uu} = \frac{1}{2} [ Q_1^2(u) + Q_2^2(u)]\,.
\end{equation}
The same is true for the Einstein tensor with its single non-vanishing component given by
\begin{equation}
    G_{uu} = \frac{1}{2}\nabla^2 w(u,x,y) = \frac{1}{2}(\partial_x^2 +\partial_y^2) w(u,x,y)\,.
\end{equation}
Note that the 4d Laplacian of $w(u,x,y)$ does not involve $u$-derivatives. In terms of complex coordinates $\zeta = (x+ i y)/\sqrt{2}$, the differential equation becomes
\begin{equation}
 \partial_\zeta\partial_{\bar{\zeta}} w(u,\zeta, \bar{\zeta}) = \frac{1}{2}[ Q_1^2(u) + Q_2^2(u)]\,, 
\end{equation}
meaning that the solution can be written as 
\begin{equation}
    w(u, \zeta, \bar{\zeta}) = f(u, \zeta) + \bar{f}(u, \bar\zeta)+\frac{1}{2} \zeta \bar{\zeta}[ Q_1^2(u) + Q_2^2(u)]\,,
\end{equation}
where $f$ is a complex function. In the absence of an electromagnetic field strength, the wave is purely gravitational. Finally, we note that the pp-wave metric is such that one obtains the same expressions when considering Born-Infeld instead of Maxwell electrodynamics, i.e. the field strength still solves BI equations, sources the same stress tensor and hence leads to the same solutions of the Einstein equations.

\subsection{Uplift of pp-waves to 5d AdS}
\label{subsec:pp_waves_uplift}

Similarly, AdS plane waves have been studied in many references before, including \cite{Cvetic:1998jf,Behrndt:1999jp,Gurses:2014soa}. The metric takes the following form:
\begin{equation}
    \dd s_5^2 = \frac{1}{k^2 z^2}\left( - 2 \dd u \dd v -w(u,x,y,z) \dd u^2 + \dd x^2 + \dd y^2 + \dd z^2 \right)\,.
\end{equation}
Here we are using coordinates $z$ such that $1/z$ is the radius of the dark bubble. This choice makes it is easier to compare with the literature on holographic renormalisation. In later sections, we will convert to $\xi=1/z$ in order to better understand the dark bubble physics. 
Note that, if $w$ vanishes, this is just the standard Poincaré metric of AdS.
We will consider solutions of the five-dimensional action
\begin{equation}
    S = \frac{1}{16\pi G_5} \int \sqrt{-g_5}\left(R_5 -3 \Lambda_5 -\frac{1}{2}(\partial \phi)^2 - \frac{1}{2}\e^{-(\phi-\phi_0)}|H|^2 \right)\,,
\end{equation}
where $\Lambda_5 = - 4 k^2$ and $H=\dd B$ is a three-form field strength. Very similarly to the 4d case, there exists a class of solutions to the $H$-equation of motion of the form
\begin{equation}
    H =\frac{1}{k z}\left[ q(u) \dd u \wedge \dd x \wedge \dd z  - p(u) \dd u \wedge \dd y \wedge \dd z - z s(u) \dd u \wedge \dd x \wedge \dd y \right]\,.
\end{equation}
It is also interesting for later to note that
\begin{equation}
    \star H = p(u) \dd u \wedge \dd x + q(u) \dd u \wedge \dd y + z s(u) \dd u \wedge \dd z\,.
\end{equation}
The equations of motion allow for a non-trivial dilaton as a function $u$, but for simplicity in this paper, we choose to freeze the dilaton $\phi-\phi_{0} = 0$. Then, similarly to the four-dimensional case,  the five-dimensional stress tensor has only one component too, which is 
\begin{equation}
    T_{uu} =  k^2 z^2 \left[p^2(u) +q^2(u) + z^2 s^2(u)\right]\,,
\end{equation}
and very similarly, the only non-trivial component of the Einstein equation is the $uu$-component:
\begin{equation}
\frac{1}{2 k^2 z^2} \nabla^2 w = \frac{1}{2} \left[ \left(\partial_z^2- 3 \frac{\partial_z}{z}\right)w + (\partial_x^2 + \partial_y^2)w \right] = \frac{1}{2} \left(k^2 z^2\left[p^2(u) +q^2(u) + z^2 s^2(u)\right]\right)\,,
\end{equation}
This is a linear PDE with a source term, and so the solution is a linear combination of a solution to the homogeneous equation and a particular solution.
A particular solution to this problem can be found as follows:
\begin{equation}\label{eq:w_particular}
    w_p(u,z) =\frac{1}{48} k^{2}z^4 \left(3 \,[q^2(u)+p(u)^2] \,(4 \log(z)-1)+4 \,z^2 s(u)^2\right)\,.
\end{equation}
The solution to the homogeneous equation is given by
\begin{equation}
    w_h(u,x,y,z) =  w_4(u,x,y) + w_g(u,z) + w_m(u,x,y,z)\,,
\end{equation}
where the individual terms satisfy
\begin{subequations} \label{eq:w_solutions}
\begin{align}
    &(\partial_x^2 + \partial_y^2)\,w_4(u,x,y) = 0\,, \label{eq:w_xy}\\
    &w_g(u,z) = c(u)\, z^4\,,\label{eq:w_uz} \\
    &w_m(u,x,y,z) = \int \dd^2 \lambda \;\e^{i \Vec{\lambda}\cdot \Vec{x}}z^2\left[A(u, \vec\lambda)\; K_2(\lambda z) + B(u,\vec\lambda )\; I_2(\lambda  z)\right]\,,\label{eq:w_uxyz}
\end{align}
\end{subequations}
with $\Vec{\lambda} = (\lambda_x , \lambda_y)$ and $\Vec{x} = (x,y)$. We also denote $\dd^2 \lambda = \dd \lambda_x \,\dd \lambda_y$ and $\lambda = ||\vec{\lambda}||$ for brevity.
To simplify the setup, we will look at configurations where all the $u$-dependence factors out. We can make this choice, as the equations of motion do not constrain the $u$-dependence. We also set the function $s(u)$ to zero, effectively setting $H_{uxy}=0$. With this, the $u$-dependent eq.~\eqref{eq:w_uz} and \eqref{eq:w_uxyz} can be written as:
\begin{subequations} \label{eq:w_solutions}
\begin{align}
    w_g(u,z) &= M^2(u)\,c\, z^4\,,\\
    w_m(u,x,y,z) &= M^2(u)\int \dd^2 \lambda \;\e^{i \Vec{\lambda}\cdot \Vec{x}}z^2\left[A( \vec\lambda)\; K_2(\lambda z) + B(\vec\lambda )\; I_2(\lambda  z)\right]\,.
\end{align}
\end{subequations}
The physical setup is also subject to boundary conditions, which determines $c$, $A$, $B$ and $w(u,x,y)$. In the final dark bubble setup that we are interested in, we glue two such spacetimes together across a brane. The boundary conditions are provided by requirements of regularity together with the junction conditions across the brane. The equations are similar to the case of massless sources in \cite{Danielsson:2025aa}, and we need to impose the further boundary condition which was argued for there. This is equivalent to the use of mixed boundary conditions at holographic infinity rather than Dirichlet ones. We will come back to this later. 
Let us now focus on the boundary conditions provided by the junction conditions.

\subsection{The junction conditions}
We glue two AdS$_{5}$ spacetimes with pp-waves  across a brane, so that we induce a pp-wave spacetime within the brane. 
To do this, we consider 5d spacetimes with different cosmological constants and different coordinates, and we will denote them by an index $-$ for the inside and $+$ for the outside. Then, we put a brane at $z_+ = z_- = z_0$.
The first junction condition tells us that the induced metric across the hypersurface at $z_\pm=z_0$ should remain continuous:
\begin{equation}\label{eq:first_junction_condition}
    \dd s_4^2|_{+} = \dd s_4^2|_{-}\,.
\end{equation}
This requires a coordinate redefinition as follows:
\begin{equation}\label{eq: scalings}
    \{u_\pm, v_\pm, x_\pm, y_\pm \} = k_\pm z_0 \{u,v,x,y\}\,, 
\end{equation}
where the coordinates $\{u,v,x,y\}$ are proper coordinates on the brane. In these variables, the first junction condition requires also
\begin{equation}\label{eq:first_junction_condition}
    w_-(u, x,y,z_0) = w_+(u,x,y,z_0) \equiv  w_\mathrm{4d}(u,x,y)\,, 
\end{equation}
such that we can write the induced metric as in eq.~\eqref{eq:4d_wave_metric}:
\begin{equation}
    \dd s^2_\mathrm{induced} = h_{mn}\dd x^m \dd x^n = -2 \dd u \dd v - w_\mathrm{4d}(u,x,y)\, \dd u^{2} + \dd x^2 + \dd y^2\,.
\end{equation}
The second junction condition reads 
\begin{equation}\label{eq:second_junction_condition}
    \left[K_{mn} - K \,h_{mn}\right]^-_+ = - \ell_5^3 S_{mn}\,,
\end{equation}
where $S_{mn}$ is the stress tensor from the brane, which is derived from the DBI action:
\begin{equation}
    S_\mathrm{brane} = - \sigma \int \dd^{4} \xi \, \sqrt{-\det(h_{mn}+ \tau \mathcal{F}_{mn})}\,,
\end{equation}
where $\sigma$ is the tension and $\tau\mathcal{F} = \tau F + B  $ is the field strength of the brane. Note that we have introduced the Planck length through $\ell_d^{d-2}=8\pi G_d$ for convenience.
The field strength is constrained by an additional junction condition, the reason being that due to its coupling to the $B$-field, the brane acts as a electric source for $H$, as its equation of motion is:
\begin{equation}
    \frac{1}{2 \ell_5^3}\dd \star_5 H =  \sigma \tau\,  \star_4 \mathcal{F} \wedge \delta(z-z_0) \dd z\,.
\end{equation}
Note that in this spacetime, these expressions are valid both for the full DBI action and its approximation to the Maxwell action. Integrating this for the relevant components of $H$, we get that
\begin{equation}
    \star_4 \mathcal{F} = \frac{1}{2 \ell_5^3 \sigma \tau \, } \left[\star_5 H\right]^{+}_{-} .
\end{equation}
We work with a configuration where there is the $H$-field is trivial on the inside, as otherwise, this would make the metric on the inside diverge at the center. With that, the field strength becomes
\begin{equation}
    \star_4 \mathcal{F} = \frac{k_+^2 z_0^2}{2 \ell_5^3\,\sigma \tau\, }\big(p_+(u) \dd u \wedge \dd x + q_+(u) \dd u \wedge \dd y\big)\,.
\end{equation}
For the pp-wave spacetime and the expression of field strength, the stress tensor of the brane is the same for both Maxwell- and Born-Infeld electrodynamics, and is given by
\begin{equation}
    S_{mn} \equiv \frac{-2}{\sqrt{-h}}\frac{\delta S_\mathrm{brane}}{\delta h^{mn}}\,, \qquad S_{mn} = - \sigma h_{mn} + S_{mn}^\mathrm{EM}\,,
\end{equation}
with the electromagnetic stress tensor given by
\begin{equation}\label{eq:em_tensor_4d}
    S_{uu}^\mathrm{EM} = \sigma \tau^2 \left(\frac{k_+^2 z_0^2}{2 \ell_5^3 \sigma \tau} \right)^2 \left(p_+^2(u) + q_+^2(u)\right)\,.
\end{equation}
The second junction condition then reduces to 
\begin{subequations} \label{eq:second_junction}
\begin{align}
 3 \ell_5^{-3} (k_- - k_+) &=\sigma\,,\\
    -\frac{1}{2}\left[k_- z_0 \partial_z w_-(u,x,y,z_0) -k_+ z_0 \partial_z w_+(u,x,y,z_0) \right] &= S_{uu}^\mathrm{EM}\,.
\end{align}
\end{subequations}
Before solving these junction conditions, we note that the second junction condition can also be recast into an Einstein equation. This was already considered in \cite{Banerjee:2020ab}, where it was argued that the Gauss-Coddazi equations in combination with the junction conditions, lead to the following equation
\begin{equation}
\label{eq:4dEinsteinEq_Exact_small_k_main}
    \left(\frac{1}{k_+} - \frac{1}{k_-}\right)G_{mn} = \left(\frac{\mathcal{J}_{mn}^+}{k_+} - \frac{\mathcal{J}_{mn}^-}{k_-}\right) -\frac{1}{2}\left(\frac{\mathcal{J}^+}{k_+} - \frac{\mathcal{J}^-}{k_-}\right)h_{mn}+ 3(k_+ - k_-)h_{mn} - 2 \ell_5^3 S_{mn}\,.
\end{equation}
This result was obtained by expanding the Gauss-Coddazi equations in small inverse AdS radius $k$ and plugging in the junction conditions in ref.~\cite{Banerjee:2019aa}. We discuss this in more detail in Appendix \ref{app:4dEinsteinEquations}. It is important to point out that in ref.~\cite{Banerjee:2019aa}, this result is only valid at leading order in $k$, whereas for the pp-wave spacetime here, eq.~\eqref{eq:4dEinsteinEq_Exact_small_k_main} is \textit{exact}, and does not require any corrections. 

\begin{figure}[h]
    \centering
    \includegraphics[width=0.75\textwidth]{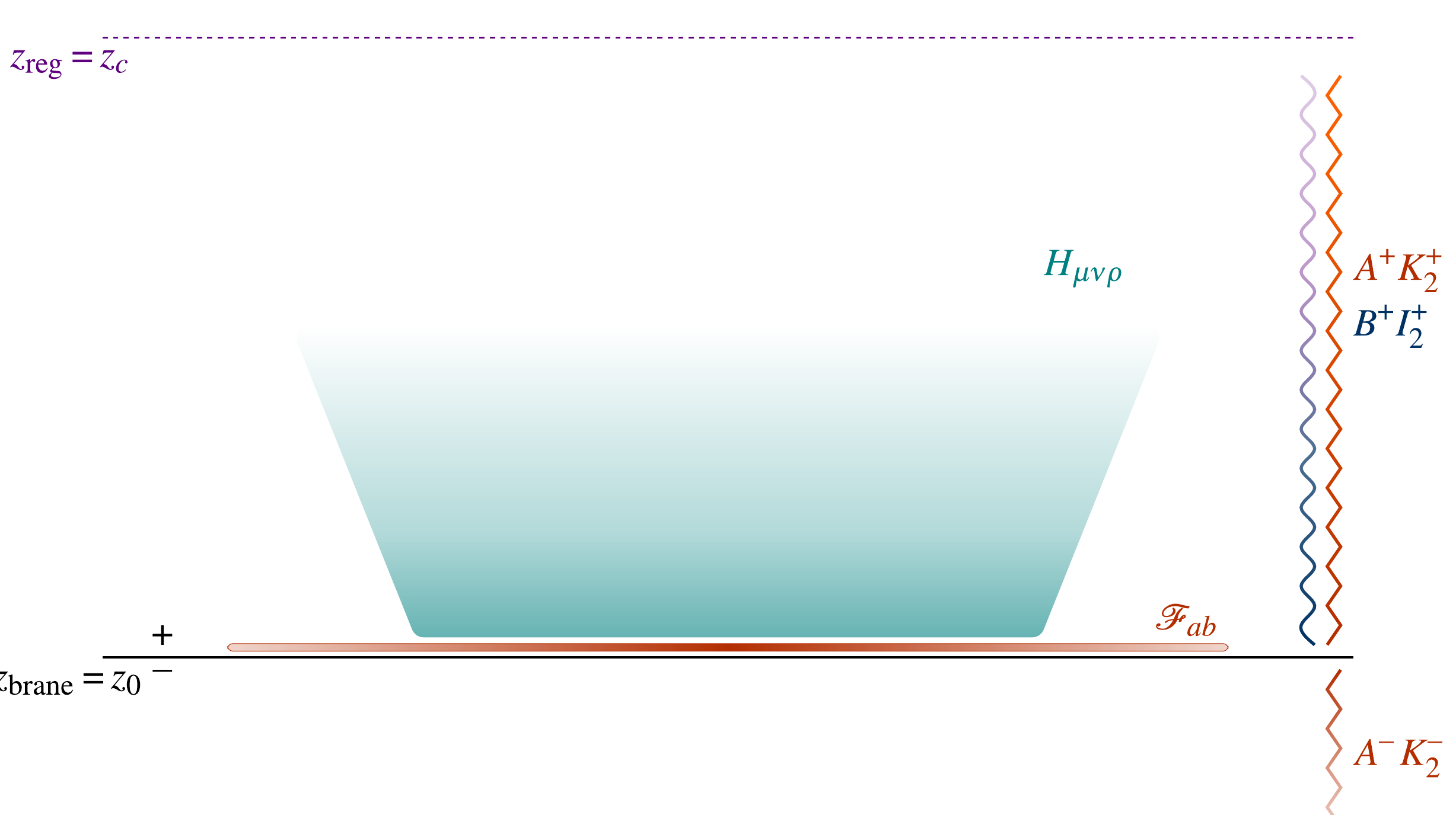}
    \caption{The figure shows the dark bubble brane, with $\mathcal{F}$ on top, which sources $H$. When the presence of these are accounted in the backreaction, we get the non-normalisable modes $K_{2}$ and normalisable modes $I_{2}$. The holographic regulator is also displayed in the figure. }
    \label{fig:yourlabel}
\end{figure}

\subsection{An additional boundary condition}
The junction conditions provide boundary conditions for our setup, but they are not sufficient. For this reason, an additional mixed boundary condition was proposed in \cite{Danielsson:2025aa}, and it boils down to requiring gravity at a holographic screen at $z= z_c$ to be  a ``rescaled'' version of the theory on the dark bubble, that is, 
\begin{equation}
\label{eq:extra_condition}
    G_{mn}^{\mathrm{(4d)}}|_{z_c} = \ell_4^2 \tau^{-1}_\mathrm{reg}T_{mn}^\mathrm{reg}|_{z_c},
\end{equation}
where $\tau_\mathrm{reg}=\tau_\mathrm{reg}(z_c^2/z_0^2)$ is a proportionality constant parametrically depending on $z_c^2/z_0^2$ that has to be fixed.\footnote{In this paper $z$ is defined such that $z= 0$ is the boundary of AdS, while in ref.~\cite{Danielsson:2025aa}, $z$ is defined so that the boundary is at $z=\infty$.} We impose the condition at small $z_c$ close to the boundary, and ignore any subleading terms. It is important to realise that the dark bubble model cannot yield a unique prediction unless supplemented by such a boundary condition. The physical meaning of our choice is that the bulk is empty and that no new structures are encountered above the dark bubble, neither as the universe expands nor through holographic scaling. 

An amusing parallel, suggested in \cite{Danielsson:2025aa}, is the non-renormalisability of quantum gravity. The problem is not that the theory is divergent containing infinities. The problem is that an infinite number of counterterms are needed, which renders the theory incapable of making any relevant predictions. The dark bubble model is similar in the sense that we might have any kind of structures in the higher dimensional throat above the dark bubble affecting physics at high energies. In that sense there is a lack of predictability. The difference is that there is a natural condition that we can impose, given by (\ref{eq:extra_condition}), which makes the model predictive. We will discuss this more precise a little bit later.

We derive the Einstein equation \eqref{eq:extra_condition} at the cut-off surface $z_c$ from the boundary action, which includes a boundary gravity term, the typical GHY boundary term and a counterterm \cite{Balasubramanian:1999re,Papadimitriou:2004ap,Papadimitriou:2004rz} as follows:
\begin{equation}
    S_\mathrm{boundary} =  S_\mathrm{bound.\ grav.} + S_\mathrm{GHY} + S_\mathrm{cntr}\,.
\end{equation}
The separate terms are given by
\begin{subequations}
    \begin{align}
    S_\mathrm{GHY} &=  \frac{1}{\ell_5^3}\int \dd^{4} \xi\, \sqrt{h} K\,,\\
    S_\mathrm{cntr} &= \frac{1}{\ell_5^3}\int \dd^{4} \xi\,\sqrt{h}
    \Bigg(
        \tilde{\kappa}_{1}R
        -\tilde{\kappa}_{2}
        + \log(z_c)
        \bigg[
            \tilde{\kappa}_{3}\Big(R_{mn}R^{mn}-\tfrac{1}{3}R^2\Big) + \tilde{\kappa}_{4}(F_H)_{mn}(F_H)^{mn}
        \bigg]
    \Bigg)\,,
\end{align}
\end{subequations}
with $\{\tilde{\kappa_{i}}\}$ as coefficients to be determined. Note that we defined the two-form $F_H = (\star_5 H)|_{\mathrm{(4d)}}$ where we pull it back on the hypersurface. More precisely, we find
\begin{equation}
    (F_H)_{ux} = (k z_c)^2 p(u)\,, \qquad (F_H)_{uy} = (k z_c)^2 q(u)\,,
\end{equation}
and all other components are vanishing.
We have to add this counterterm, as the appearance of the $1/z$ in the solution of $H$ renders the on-shell action to be log-divergent.
Additionally, we understand $G_{mn}^{\mathrm{(4d)}}|_{z_c}$ as a boundary gravity \cite{Compere:2008us}.
We get, as before,
\begin{equation}
    G_{uu}^{\mathrm{(4d)}}|_{z_c} = \frac{1}{2} \nabla_\mathrm{(4d)}^2 w_+(u,x,y,z_c).
\end{equation}
For the regularised stress tensor, we then get \cite{Balasubramanian:1999re,Papadimitriou:2004ap,Papadimitriou:2004rz,Bernamonti:2007bu}:
\begin{equation}
    \ell_{5}^{3}\,T^{\rm reg}_{mn}\big|_{z_{c}} = T^{K}_{mn}\big|_{z_{c}} + \left\{ \left(\tilde{\kappa}_3\,T^{R}_{mn}+ \tilde{\kappa}_{4}\,T^{F}_{mn}\right)\log(z)\right\}\big|_{z_{c}}\, 
\end{equation}
where
\begin{alignat}{1}
    T^{K}_{mn} &= K_{mn} - K\,h_{mn} -2  \tilde{\kappa}_1 \,G_{mn}^\mathrm{(4d)} + \tilde{\kappa}_{2}\, h_{mn}\,,\nonumber\\
    T^{R}_{mn} &= h_{mn}\,\left(R_{ij}R^{ij} -\tfrac{1}{3}R^2\right) +4 R_{minj}R^{ij}-\tfrac{4}{3}R R_{mn}-2 \nabla^2 G^{(4d)}_{mn} +\tfrac{2}{3}\left(\nabla_m\nabla_n
    -h_{mn}\nabla^2\right)R\,,\\
    T^{F}_{mn} &= h_{mn}\, (F_H)_{ij}(F_H)^{ij}-4(F_H)_{mi}(F_H)^{i}_{\;\;n}\,\nonumber.
\end{alignat}
The Ricci scalar vanishes for the induced metric, and so do the combinations $R_{ij}R^{ij}$ and $R_{minj}R^{ij}$, such that only the term with the Laplacian acting on the Einstein tensor contributes to the gravitational logarithmic terms. The norm of $F_H$ vanishes, so only the second term in $T^{F_H}_{mn}$ contributes.
The condition \eqref{eq:extra_condition} can then be rewritten as
\begin{equation}
    \begin{aligned}
    \left[\frac{\ell_5^3}{\ell_4^2}\tau_\mathrm{reg}+2\left(\tilde{\kappa}_1+ \tilde{\kappa}_3\log(z_c)\nabla^2_\mathrm{(4d)} \right) \right] G_{mn}^\mathrm{(4d)}\big|_{z_c} = (K_{mn} &- K h_{mn}) + \tilde{\kappa}_2 \,h_{mn}\\
    &-4\,\ell_5^3\,\tilde{\kappa}_{4}\,\log(z_c)\,(F_H)_{mi}(F_H)^{i}{}_{n}\,.
    \end{aligned}
\end{equation}
For the condition to be solved, we must choose
\begin{equation}
    \tilde{\kappa}_2 = -3 k_+\,,
\end{equation}
which ensures that the terms proportional to the induced metric (of which one is hiding in $K_{mn}$) vanish. Then only the $uu$-component remains.
It turns out that the contributions in all the quantities coming from the particular solution $w_p(u,z)$ cancel against the logarithmic term coming from the $F_H$ counterterm, if one chooses
\begin{equation}
    \tilde{\kappa}_4 = \frac{1}{8 k_+}\,.
\end{equation}
The condition then becomes 
\begin{equation}\label{eq:extra_condition_PDE}
    \left[\frac{\ell_5^3}{\ell_4^2}\tau_\mathrm{reg}+2\left(\tilde{\kappa}_1+ \tilde{\kappa}_3\log(z_c)\nabla^2_\mathrm{(4d)} \right) \right] \nabla^2_\mathrm{(4d)} w_h(u,x,y,z)\big|_{z_c} = - k\, z_c \,\partial_z w_h(u,x,y,z)\big|_{z_c}\,.
\end{equation}
We solve this by plugging in the expression for $w_h$ and pick $c_+=0$. In 4d coordinates adopted to the $z=z_c$ slice (with $z_c$ small), we can write
\begin{equation}
    \begin{aligned}
    w_h(u,x,y,z) = w_{4+}(u,x,y) + M^2(u) \int \dd^2 \lambda_+ &\,\e^{i (k_+ z_c) \vec{\lambda}_+\cdot \vec{x}}\, 
    \times\,\\
    &z^2\left[A_+( \vec\lambda_+)\; K_2(\lambda_+ z) + B_+(\vec\lambda_+ )\; I_2(\lambda_+  z)\right]\,.
    \end{aligned}
\end{equation}
Note that the the function $w_{4+}(u,x,y)$ does not participate in this condition as it does not depend on $z$ and because it is a harmonic function in the four-dimensional space. Plugging this into eq.~\eqref{eq:extra_condition_PDE}, we find
\begin{equation}
    \begin{aligned}
    \left[-A_+( \vec\lambda_+)\; K_1(\lambda_+ z_c) + B_+(\vec\lambda_+ )\; I_1(\lambda_+  z_c)\right]=\,\,\,&k_+\,z_c\,\lambda_+^2\,\\
    &\times\left[\frac{\ell_5^3}{\ell_4^2}\tau_\mathrm{reg}-2\left(\tilde{\kappa}_1+ \tilde{\kappa}_3\log(z_c)(k_+\, z_c \,\lambda_+)^{2} \right) \right]\,\\  
    &\times\left[A_+( \vec\lambda_+)\, K_2(\lambda_+ z_c) \,\,+ B_+(\vec\lambda_+ )\, I_2(\lambda_+  z_c)\right] .\\
    \end{aligned}
\end{equation}
This is a linear equation in the coefficients $A_+$ and $B_+$ and hence they can be expressed in terms of each other. However, let us refrain from doing that for now, and expand the equation in small $z_c$, keeping track of the $z_c^2/z_0^2$-dependence of $\tau_{\rm reg}=\tau_{\rm reg}( z_c^2/z_0^2)$, as well as of $\tilde{\kappa}_1 =\tilde{\kappa}_1( z_c^2/z_0^2)$. Expanding the result, order by order in $z_{c}$ and ignoring $O(z^6)$ contributions, we find:
\begin{equation}
    0=- A_+(\vec\lambda_+ ) \;C \; k_+^2 z_c^2 + \frac{1}{2}k_+ \lambda_+^2 z_c^4\biggl[ B_+(\vec\lambda_+ )-\eta(\lambda_+) A_+(\vec\lambda_+ ) \biggr] + O(z_c^6)\,,
\end{equation}
where we defined $C$ and $\eta(\lambda_+)$ as follows: 
\begin{align}
    &C = k_+^{-1} + 4 \tilde{\kappa}_1 (0) + 2 \frac{\ell_5^3}{\ell_4^2} \tau_\mathrm{reg}(0)\,,\\
    &\eta(\lambda_+) = \biggl(\gamma_\mathrm{EG}-\frac{k_+C}{2} - (4 k_+^3 \tilde{\kappa}_3 - 1) \log(z_c) +  \log\left(\frac{\lambda_+}{2}\right)
    +\frac{4 k_+}{z_0^2\lambda_+^2}\left[2 \tilde{\kappa}_1'(0)+ \frac{\ell_5^3}{\ell_4^2} \tau_\mathrm{reg}'(0)\right]\biggr)\,.
\end{align}
We then choose $C =0$, such that the leading term vanishes, and we choose $\tilde \kappa_3 = 1/(4k_+^3)$, such that the logarithmic contribution in $z_c$ in $\eta(\lambda_+)$ vanishes. Then, at order $z_c^4$, we find that 
\begin{equation}
\label{eq: extra_cond}
    B_+(\vec\lambda_+ )= A_+(\vec\lambda_+ ) \left[\frac{4 \,k_+}{z_0^2\,\lambda_+^2}\left[2 \,\tilde{\kappa}_1'(0)+ \frac{\ell_5^3}{\ell_4^2}\, \tau_\mathrm{reg}'(0)\right]  + \gamma_\mathrm{EG} + \log(\lambda_+/2) \right]  .
\end{equation}
Keeping only the leading pieces in the momentum $\lambda_+$, and choosing
\begin{equation}
    \tau_\mathrm{reg}(0) = 0, \qquad \tau_\mathrm{reg}'(0) = 1, \qquad \tilde{\kappa}_1 (0) = -\frac{1}{4\,k_+}, \qquad \tilde{\kappa}_1'(0)=\frac{\ell_5^3}{\ell_4^2}\frac{\Delta\kappa_1}{2}\,,
\end{equation}
this reduces to:
\begin{equation}
\label{eq: extra_cond_new}
    B_+(\vec\lambda_+ )= A_+(\vec\lambda_+ ) \frac{4 \,k_+(1+\Delta \kappa_1)}{z_0^2\,\lambda_+^2} \frac{\ell_5^3}{\ell_4^2}\,.
\end{equation}
If we choose $\Delta \kappa_1=1$, it is the same mixed boundary condition as obtained in \cite{Danielsson:2025aa}, relating the non-normalisable modes (proportional to $A_+$) to the normalisable modes (proportional to $B_+$). \footnote{The choice $\Delta \kappa_1=1$ guarantees that $G_4$ is correctly defined. }

The choice of dropping higher order terms in $z_c$ in (\ref{eq:extra_condition}) is equivalent to there being no subleading contributions in the momentum. That is, (\ref{eq: extra_cond_new}) is supposed to be {\it exact}. This is easier to understand using the coordinates used in \cite{Danielsson:2025aa}. As we move closer to the cut-off, holding the proper momentum on the brane fixed, the momentum at the cut-off is redshifted towards zero and no further dependence is picked up. If you start with too high proper momentum on the brane, and the cut-off not sufficiently close to the boundary, subleading terms would hit back. This is the familiar from cut-off regularisation, where the UV cut-off needs to sit an energy scale much larger than the energy scales of interest. We emphasise again that the cut-off does not correspond to a physical brane but is just a way to control the boundary conditions.

As explained in \cite{Danielsson:2025aa}, and reviewed earlier, one can in principle choose any boundary condition consistent with the rest of the system. Just as in the case of AdS/CFT, the condition one chooses is dictated by what kind of physics that the system is supposed to describe. We have argued that the condition we choose is unique within the natural physical assumptions that we have specified. 

\subsection{Combining the boundary conditions}
Having discussed the origin of some of the boundary conditions, we now proceed translating this into conditions on the functions $w_\pm (u,x,y,z)$ on both the in- and outside. 
If we require the solution to remain finite in the interior, i.e.
\begin{equation}
    \lim_{z \to +\infty} w_-(u,x,y,z) = w_-(u,x,y,\infty) < \infty\,,
\end{equation}
then we must set
\begin{equation}
    c = 0\,, \qquad B(\vec{\lambda}) = 0\,, \qquad w_p(u,z)=0\,, \qquad w_{4-}(u,x,y) = w_-(u,x,y,\infty)\,.
\end{equation}
By regularity, we require the $H$-field to be trivial in the interior.
Implementing these boundary conditions, the solution on the inside looks as follows:
\begin{equation}
    w_-(u,x,y,z) = w_{4-}(u,x,y) + M^2(u) \int \dd^2 \lambda \e^{i \vec{\lambda} \cdot \vec{x}} \frac{z^2}{(k_- z_0)^2} A_-(\vec{\lambda}) K_2\left(\frac{\lambda z}{k_- z_0}\right)\,.
\end{equation}
Note that we have performed the coordinate transformation $\{x_-,y_-\} = k_- z_0 \{x, y\}$  and redefined $\vec{\lambda}_- = \vec{\lambda}/k_- z_0$. The $z$-derivative of this function, relevant for the second junction condition, gives
\begin{equation}
    \partial_z w_-(u,x,y,z) =  -M^2(u) \int \dd^2 \lambda \e^{i \vec{\lambda} \cdot \vec{x}}\frac{\lambda z^2}{(k_- z_0)^3} A_-(\vec{\lambda}) K_1\left(\frac{\lambda z}{k_- z_0}\right)\,.
\end{equation}
On the outside, where the $H$-field is turned on, the solution takes the full form
\begin{multline}
    w_+(u,x,y,z) = w_p(u,z)+w_{4+}(u,x,y) + M^2(u) c z^4\\
    + M^2(u) \int \dd^2 \lambda \e^{i \vec{\lambda} \cdot \vec{x}} \frac{z^2}{(k_+ z_0)^2} \left[ A_+(\vec{\lambda}) K_2\left(\frac{\lambda z}{k_+ z_0}\right)+B_+(\vec{\lambda}) I_2\left(\frac{\lambda z}{k_+ z_0}\right) \right]\,.
\end{multline}
The $z$-derivative of this function is given by
\begin{multline}
    \partial_z w_+(u,x,y,z) = \partial_z w_p(u,z)+ 4 M^2(u) c z^3\\
    + M^2(u) \int \dd^2 \lambda \e^{i \vec{\lambda} \cdot \vec{x}} \frac{\lambda z^2}{(k_+ z_0)^3} \left[ -A_+(\vec{\lambda}) K_1\left(\frac{\lambda z}{k_+ z_0}\right)+B_+(\vec{\lambda}) I_1\left(\frac{\lambda z}{k_+ z_0}\right) \right]\,.
\end{multline}
The first junction condition \eqref{eq:first_junction_condition} then gives
\begin{multline}
    w_{4-}(u,x,y) - w_{4+}(u,x,y) - w_p(u,z_0) - M^2(u) c z_0^4\\ = M^2(u)  \int \dd^2 \lambda \e^{i \vec{\lambda} \cdot \vec{x}} \left[\frac{A_+(\vec{\lambda})}{k_+^2} K_2\left(\frac{\lambda }{k_+}\right)+\frac{B_+(\vec{\lambda})}{k_+^2} I_2\left(\frac{\lambda}{k_+}\right)-\frac{A_-(\vec{\lambda})}{k_-^2}K_2\left(\frac{\lambda }{k_-}\right) \right]\,.
\end{multline}
We can write the left hand side in terms of a Fourier expansion as well. For that, we take $p_+(u) = p_+ M(u)$ and $q_+(u) = q_+ M(u) $ and we will put $s(u)=0$, such that the $u$-dependence in eq.~\eqref{eq:w_particular} can be factored out and is the same as in the right hand side. With that, we write, 
\begin{equation}
    w_{4-}(u,x,y) - w_{4+}(u,x,y) - w_p(u,z_0) - M^2(u) c z_0^4 = M^2(u)  \int \dd^2 \lambda \e^{i \vec{\lambda} \cdot \vec{x}} \;j( \vec{\lambda}) \,.
\end{equation}
Hence, the first junction condition becomes a condition for the Fourier coefficients:
\begin{equation}\label{eq:first_junction_j}
    j(\vec{\lambda})=\frac{A_+(\vec{\lambda})}{k_+^2} K_2\left(\frac{\lambda }{k_+}\right)+\frac{B_+(\vec{\lambda})}{k_+^2} I_2\left(\frac{\lambda}{k_+}\right)-\frac{A_-(\vec{\lambda})}{k_-^2}K_2\left(\frac{\lambda }{k_-}\right)\,.
\end{equation}
Similarly, the second junction condition \eqref{eq:second_junction} is given by
\begin{multline}
     -\frac{1}{2}k_+ z_0 \partial_z w_p(u,z)|_{z_0} - 2 k_+ M^2(u) c z_0^4 + S_{uu}^\mathrm{EM}\\
     =-\frac{1}{2} M^2(u)\int \dd^2 \lambda \e^{i \vec{\lambda} \cdot \vec{x}} \lambda \left[\frac{A_+(\vec{\lambda})}{k_+^2} K_1\left(\frac{\lambda}{k_+}\right)-\frac{B_+(\vec{\lambda})}{k_+^2} I_1\left(\frac{\lambda}{k_+ }\right)-\frac{A_-(\vec{\lambda})}{k_-^2} K_1\left(\frac{\lambda}{k_-}\right) \right]\,.
\end{multline}
Also here, we can also Fourier transform the left-hand side as 
\begin{equation}
    -\frac{1}{2}k_+ z_0 \partial_z w_p(u,z)|_{z_0} - 2 k_+ M^2(u) c z_0^4 + S_{uu}^\mathrm{EM} = -\frac{1}{2}M^2(u) \int \dd^2 \lambda \e^{i \vec{\lambda} \cdot \vec{x}} \; T( \vec{\lambda})\,.
\end{equation}
Hence, the second junction condition becomes 
\begin{equation}\label{eq:second_junction_T}
    T( \vec{\lambda}) = \frac{A_+(\vec{\lambda})}{k_+^2} K_1\left(\frac{\lambda}{k_+}\right)-\frac{B_+(\vec{\lambda})}{k_+^2} I_1\left(\frac{\lambda}{k_+ }\right)-\frac{A_-(\vec{\lambda})}{k_-^2} K_1\left(\frac{\lambda}{k_-}\right)\,.
\end{equation}
Finally, we use the extra condition~\eqref{eq: extra_cond_new}, writing it as $B_+(\vec{\lambda}) = \eta A_+(\vec{\lambda}) / \lambda^2$. Then, we invert eqns.~\eqref{eq:first_junction_j} and \eqref{eq:second_junction_T} to solve for $A_+(\vec{\lambda})$ and $A_-(\vec{\lambda})$:
\begin{equation}
\begin{pmatrix}
    A_+ \\
    A_-
\end{pmatrix}
= D \,
\begin{pmatrix}
  k_+^2\, I_{1+} \,\lambda^2 \,(j\, K_{1-} - T\, K_{2-} )\\ 
  -k_-^2 [j\, (\eta - I_{1+} K_{1+} \,\lambda^2) + 
   T\, I_{1+} \,(I_{2+} \,\eta + K_{2+} \,\lambda^2)]
\end{pmatrix}\,,
\end{equation}
where
\begin{equation}
D = \left(I_{1+}\left(\eta \, K_{1-}
   I_{2+}+\lambda ^2 \,(K_{1-}
   K_{2+}-K_{2-} K_{1+})\right)+\eta\,  K_{2-} \right)^{-1}\,,
\end{equation}
Here, we have used the short notation $K_{2\pm} = K_2(\lambda/k_\pm)$ etc.

In the case where we turn off the $H$-field and remove all $z$-dependence altogether, we just find that $w$ is only determined by $w_{4\pm}$ and the junction conditions reduce to
\begin{equation}
    w_{4-}(u,x,y) = w_{4+}(u,x,y)\,.
\end{equation}
Note that this corresponds to a gravitational pp-wave on the brane and fits with the results of \cite{Danielsson:2022aa}.

\section{Wave phenomenology on the dark bubble}
\label{sec:phenomenology}

Let us now make contact with ref.~\cite{Basile:2023aa}, where electromagnetic waves on the dark bubble were first studied. In this work, the expansion of the bubble in the presence of a cosmological constant was taken into account, but the computations were limited to taking an average of waves moving in different directions and over wave lengths. The logarithmic behavior of the particular solution $w_p (u,z)$ was identified, while the homogeneous piece $w_h(u,z)$ showed up only as a constant of integration. 

As explained in \cite{Basile:2023aa}, the logarithm reflects how the mass up to a certain radius increases due to the energy density from the $B$-field. It was also noted that $w_p(u,z)$ is suppressed compared to $w_h(u,z)$. It is easy to understand why. The homogeneous piece is sourced by matter on the brane through the second junction condition. When $\Delta k=k_--k_+$ becomes small compared to $k_\pm$, the back reaction in the bulk becomes large. As already mentioned, this is reflected by Newton's constant in 4d becoming large, $G_4 \sim G_5 k^2/\Delta k$. The particular solution, on the other hand, is directly sourced by the $B$-field and lacks this factor of $k/\Delta k$.

We also note the presence of a further homogeneous piece $w(u,x,y)$ that does not depend on $\xi$.\footnote{We will in the following express everything in terms of $\xi=1/z$, which is the radius of the dark bubble. }  This is nothing else than the gravitational waves studied in \cite{Danielsson:2022aa}. In that paper no assumption of  pp-waves were made, and higher order corrections are expected.

We will find a structure very similar to \cite{Danielsson:2025aa} for point masses. As explained in that work, there are essentially two ways for gravity to get modified at small length scales. 

\begin{enumerate}
    \item Gravity gets weaker in vacuum at distances smaller than $L$ from localised mass.  
    \item Gravity gets weaker at energy densities of order string scale, preventing the Hubble radius to get smaller than $L$. 
\end{enumerate}

As we will see, in the case of pp-waves, the second type of correction is absent due to a fine tuned cancellation. The pp-waves are exact solutions and waves moving in the same direction can be superimposed, despite gravity being non-linear. If pp-waves moving in different directions are added, non-linear effects bring the correction back in.
Let us now consider the pp-waves with a cylindrical
cross section so that they depend on a radial coordinate $r$. We then have
\begin{equation}
    \omega(u,r,\xi)=U(u)\,R(r)\,Z(\xi),
\end{equation}
where we now use $\xi=1/z$.
The function $U(u)$ represents the arbitrary shape of the wave in the longitudinal direction, while $R(r)$ describes the cross-section. To make it physically more explicit, we will write the complex Fourier transforms of the previous section in terms of Hankel transforms:
\begin{equation}
   \frac{1}{2\pi} \int d^2\lambda \,e^{-ip\lambda}\,  f(|\lambda|)=\int_0^\infty dr\,J_0(pr)\,r\,f(r)\,,
\end{equation}
where we have insisted that the function should be regular as $r\rightarrow 0$ so that we can drop any contribution from $Y_0(pr)$. The momentum $p$ can in principle also be a function of $u$, but let us for for simplicity focus on waves for which this is not the case, that is we only consider waves that have the same shape at all radial scales. Since the overall shape is just a multiplicative factor we drop that as well. Hence, the solution can be written
\begin{equation}
    w(r,\xi)=J_0(pr)\,\frac{1}{\xi^2}\,\left(A(p)K_2\,\left(\frac{p}{k^2\xi}\right)+B(p)\,I_2\left(\frac{p}{k^2\xi}\right)\right)\,,
\end{equation}
We now allow for different solutions on the inside and the outside of the dark bubble. We demand regularity at the center of the bubble. This gives:
\begin{equation}
    \begin{split}
        w_-(r,\xi) &= J_0(p_-r_-)\,\frac{1}{\xi^2}\,A_-(p_-)\,K_2\,\left(\frac{p_-}{k_-^2\xi}\right)\,,\\
        w_+(r,\xi) &= J_0(p_+r_+)\,\frac{1}{\xi^2}\,\left(A_+(p_+)\,K_2\left(\frac{p_+}{k_+^2\xi}\right)+B(p)\,I_2\left(\frac{p_+}{k_+^2\xi}\right)\right)\,.
    \end{split}
\end{equation}
On a brane at $\xi=\xi_0$ we need to solve the first junction condition. This requires
\begin{equation}
    \rho=k_+\xi_0r_+=k_-\xi_0r_-\,,
\end{equation}
for the proper length, and
\begin{equation}
    q=\frac{p_+}{k_+\xi_0}=\frac{p_-}{k_-\xi_0}\,,
\end{equation}
for the proper momentum on the brane. Furthermore, we need to satisfy
\begin{equation}
    A_+(q)\,K_2^{+}+B(q)\,I_2^{+}=A_-(q)\,K_2^-\,,
\end{equation}
where we have defined $K_2^\pm=K_2\left(\frac{q}{k_\pm}\right)$ and $I_2=I_2\left(\frac{q}{k_+}\right)$.
We get the induced 4d Einstein tensor
\begin{equation}
    G^v_v=\frac{1}{2k_\pm \xi_0^2}\,\left(\partial_r^2+\frac{1}{r}\partial_r\right)\,w(r,\xi)=-\frac{q^2}{2\xi_0^2}\,J_{0}(q\rho)\,A_-(q)\,K_2^{-}\,,
\end{equation}
where we have chosen to express the Einstein tensor using inside-quantities. All other components zero.
Writing down the second junction condition using the previous results, gives
\begin{equation}
    \frac{q}{2}\,J_0(qp)\,\left(B(q)\,I_1+A_-(q)\,K_1^-+A_+(q)\,K_1^-\right)=8\pi\,G_5 \,M(q)\,.
\end{equation}
As argued in \cite{Danielsson:2025aa}, we need one more condition to make sure that we get the right physics at large distances in 4d. This is (\ref{eq: extra_cond}):
\begin{equation}
    B(q)=\frac{4 k^2_+ \Delta k}{k_-q^2}\,A_+(q)=\frac{ 8 k_+^3G_5,}{G_4q^2}\,A_+(q)\,.
\end{equation}
Solving for $A_\pm(q)$ and $B(q)$, following \cite{Danielsson:2025aa}, we find
\begin{equation}
    G^u_v=8\pi G_4\,{\cal G}(q)\, M(q),
\end{equation}
with
\begin{equation}
    \begin{split}
        {\cal G}(q)&=\frac{kG_5}{G_4}\dfrac{q^2(\frac{8k^3G_5}{G_4} I_2+q^2K_2^+)K_2^-}{k q\left(\frac{8k^3G_5}{G_4} (I_2 K_1^-+I_1K_2^+)+q^2(K_1^-K_2^+-K_1^+K_2^-)\right)}  \\
        &= \frac{ q^4K_2^2}{ 8k^4 +2q^3[q (K_2^2 -K_1^2)- 3 k K_1 K_2]} + {\cal O}(\Delta k).
    \end{split}
\end{equation}
That is, we reproduce the general result
\begin{equation} \label{eq:darkmetric}
     h_{ab}=-\frac{16 \pi G_4}{q^2}{\cal G}(q)\left( T_{ab}-\frac{1}{2} T \eta_{ab} \right),
\end{equation}
but now evaluated on the pp-waves. We recall from \cite{Danielsson:2025aa} that ${\cal G}\rightarrow 1$ when $q\rightarrow 0$, and ${\cal G}(q) \rightarrow \frac{qG_5}{G_4}$ when $q\rightarrow \infty$.

\begin{figure}[h!]
    \centering
    \begin{subfigure}[b]{0.47\textwidth}
        \centering
        \includegraphics[width=\textwidth]{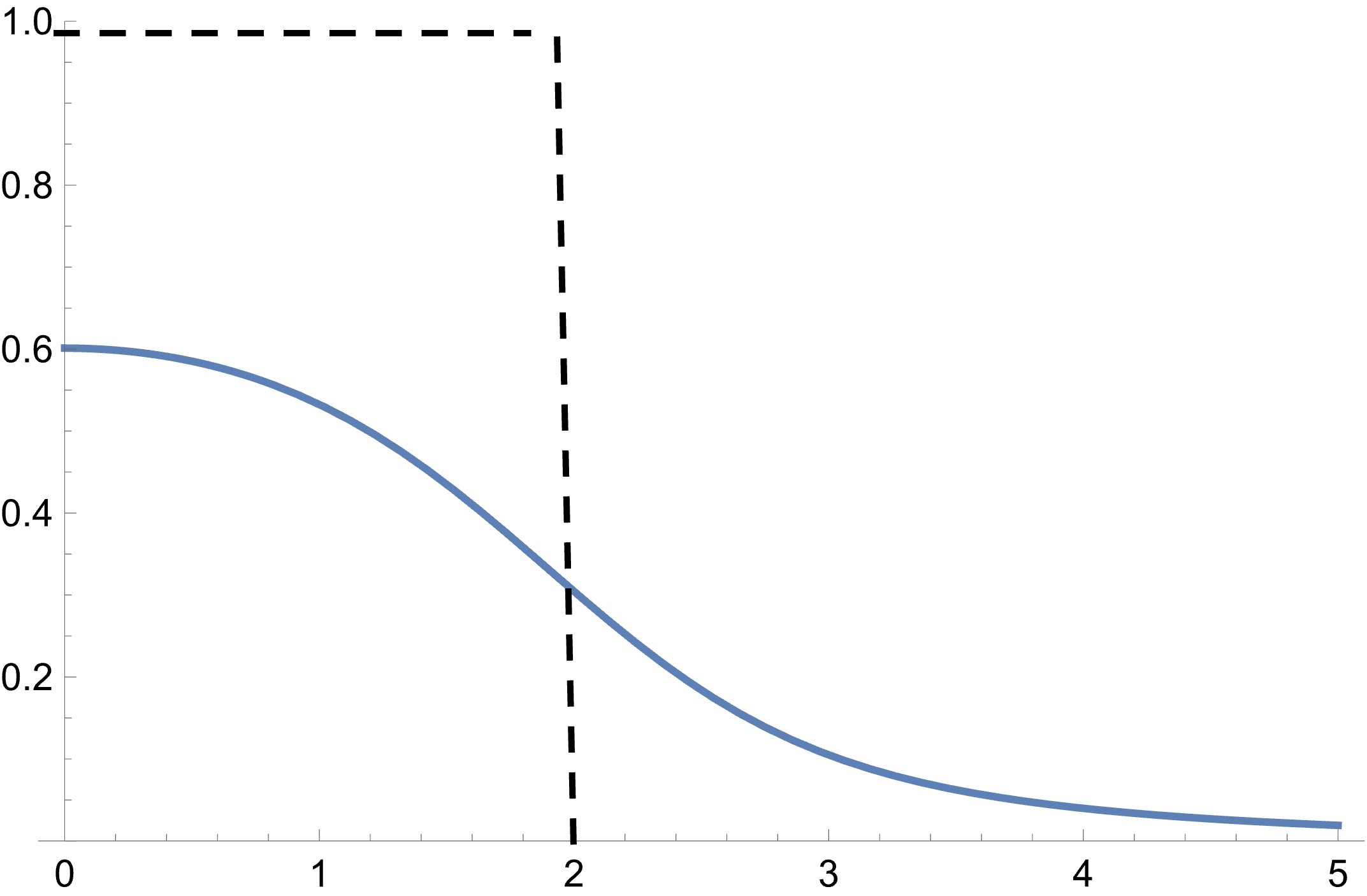}
        \caption{Broadening of a narrow beam with $a=2L$.}
        \label{fig:narrow}
    \end{subfigure}
    \hfill
    \begin{subfigure}[b]{0.47\textwidth}
        \centering
        \includegraphics[width=\textwidth]{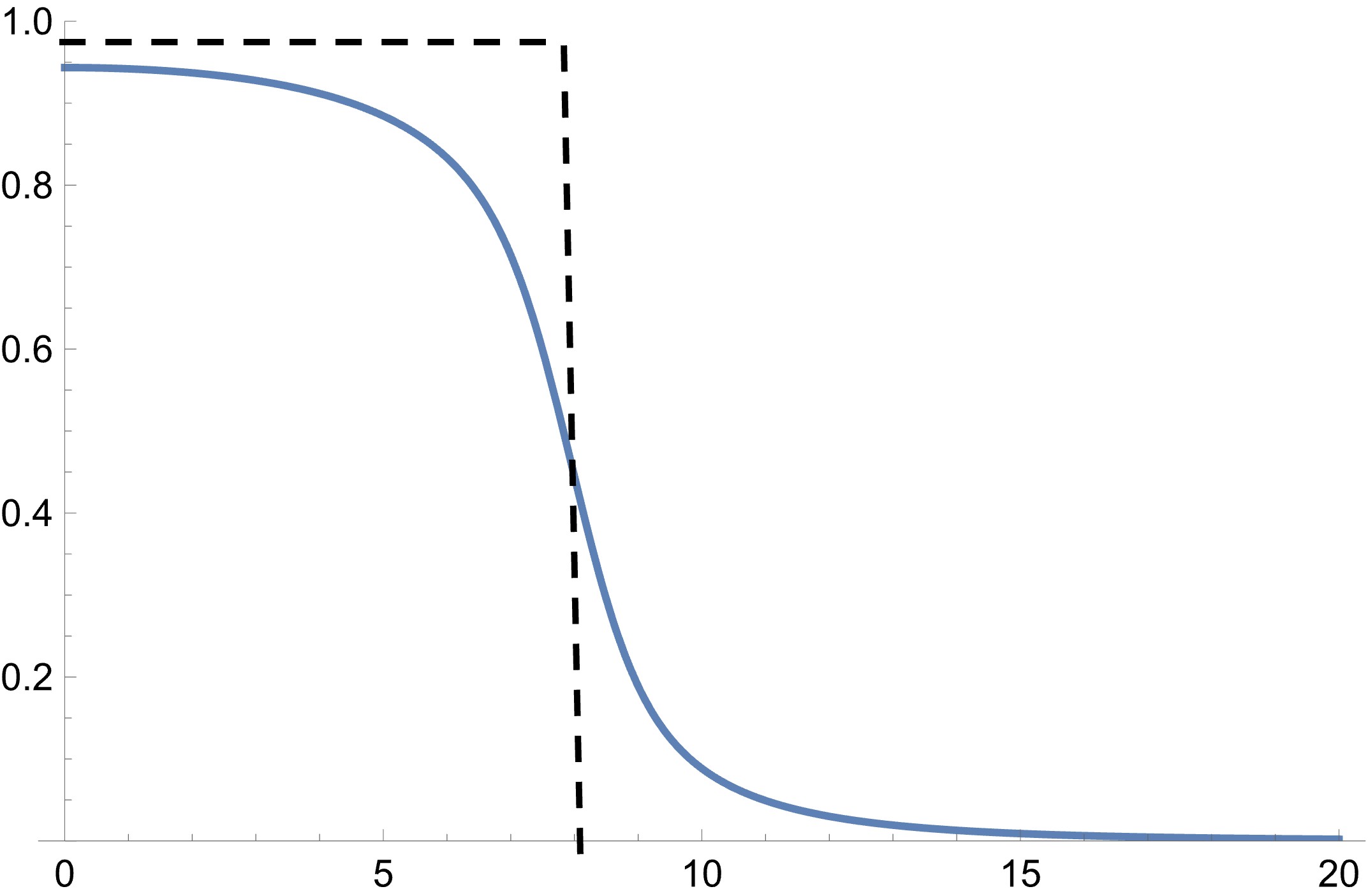}
        \caption{Broadening of a broad beam with $a=8L$.}
        \label{fig:broad}
    \end{subfigure}
\end{figure}

To be concrete, let us consider a beam of light with an energy density that is constant up to a radius $a$ where it abruptly drops to zero. The gravitational back reaction will be smeared due to gravity's inability to resolve structures smaller than $L$.
The Hankel transform of this energy density is given by
\begin{equation}
    q\, F(q)= \frac{a\,J_1(qa)}{q}\,.
\end{equation}
This leads to a modified energy density given by
\begin{equation}
    R(r)=\int_0^\infty \frac{a\,J_1(qa)}{q}\,{\cal G}(q) \,J_0(qr)\,dq\,.
\end{equation}
This is the effective energy density as seen by gravity.
Evaluating the Hankel transform numerically, we see in figure (\ref{fig:narrow})  how a narrow beam with $a=2L$ is affected much more than a broad one with $a=8L$ in figure (\ref{fig:broad}). 

\section{Discussion}
\label{sec:conclusion}

In this work we embedded pp-waves on the dark bubble by gluing two AdS$_5$ pp-wave spacetimes together across a three-brane. This provides a setting in which gravitational pp-waves on the brane are sourced by bulk gravitational waves, and electromagnetic waves on the brane are sourced by non-trivial bulk $H$-field profiles, in close analogy with earlier constructions involving electromagnetism \cite{Danielsson:2022aa,Basile:2023aa, Danielsson:2024frw}. A notable feature of the setup is that the bulk equations of motion can be solved formally exactly, and the junction conditions across the brane can also be implemented exactly. In addition, the construction requires an extra mixed boundary condition at the holographic cutoff, which fixes the relation between normalisable and non-normalisable modes and thereby completes the definition of the effective four-dimensional theory. Together, these ingredients allow us to extract the gravitational backreaction of localised wave profiles and to see how short-distance gravitational corrections emerge in the dark bubble framework as in \cite{Danielsson:2025aa}.

 The following picture emerges. The dark bubble model generates a gravitational force at large length scales, while at sufficiently small scales it essentially disappears. Einstein gravity in 4d is induced through an unconventional embedding into higher dimensions, with 4d gravity being much {\it stronger} than gravity in 5d. At length scales of order $L$ and below, gravity is universally corrected with an effective energy momentum tensor as seen by gravity (and equal to the Einstein tensor) that differs from the energy momentum tensor directly present the brane. 
 
 The coupling to bulk fields, like the coupling of the electromagnetic field to the $B$-field, is heavily suppressed and becomes relevant only at high energy. This solves a problem raised in \cite{Basile:2025aa}, where it was argued that the way non-abelian gauge fields may couple to the bulk $B$-field, could lead to a breakdown of the equivalence principle. Our results show that such effects will typically not be present. Given this, it would be 
 interesting to consider other sectors of the standard model as well, including non-abelian gauge fields, but the possibly most interesting case could be neutrinos. Since our scale $L$ is of order the dark dimension, it is likely that the neutrino sector might serve as a bridge between the standard model on the dark bubble and gravity in the bulk.

\section*{Acknowledgements} 
We would like to thank Suvendu Giri for discussions. This work was supported in part by Kungliga Fysiografiska sällskapet i Lund. The work of D.P. was supported by JSPS Kakenhi Grant-in-Aid for Scientific Research (No. 24KF0232).

\appendix
\section{The 4d Einstein equations}\label{app:4dEinsteinEquations}
One can compute the four-dimensional Einstein tensor through the Gauss-Codazzi equation, given by
\begin{equation}
    \mathcal{J}_{mn} = R_{mn} +(K_{mp}K^{p}_n - K K_{mn})\,, \qquad \mathcal{J}_{mn} =e_m^\beta e_n^\delta\left(R_{\beta \delta}^{(5)}-R^{(5)}_{\mu \beta \nu \delta} n^\mu n^\nu\right)
\end{equation}
If one does so, after going to proper coordinates on the brane, one obtains
\begin{equation}\label{eq:4dEinsteinTensor}
    G_{uu} = \frac{1}{2}\nabla^2_\mathrm{4d} w(u,x,y,z_0)\,, \qquad \mathrm{rest} = 0.
\end{equation}
which is precisely the Einstein tensor of a 4d pp wave spacetime.
To see how the stress tensor on the brane comes in, one needs to compare both inside/outside regions and use the junction conditions. They are, as usual
\begin{equation}\label{eq: 2nd_junction}
    \left[K_{mn} - K h_{mn}\right]^{+}_{-} = + \ell_5^3 S_{mn}\,,
\end{equation}
However, it is useful to trace-reverse this. Taking the trace, we note that
\begin{equation}
   3 (K^- - K^+) = + \ell_5^3 S\,.
\end{equation}
Plugging this back in eq.~\eqref{eq: 2nd_junction}, one finds
\begin{equation}\label{eq:trace_reversed}
    K_{ab}^+ -K_{ab}^- = + \ell_5^3 \left(S_{ab} - \frac{S}{3}h_{ab}\right) \equiv + \ell_5^3 S^\mathrm{TR}_{ab}\,,
\end{equation}
where $S^{\mathrm{TR}}$ represent the trace reversed contribution of the induced energy momentum tensor. This will result of great usefulness. In fact, what is special about the solution at hand, is that
\begin{equation}
    K = -4 k\,, \qquad K_{ab}K^{ab} = 4 k^2\,.
\end{equation}
Moreover, it turns out that an unique property of this background is given by:
\begin{equation}
    K_{ac} K^c_b = \frac{7}{16} K K_{ab} + \frac{1}{16} K (K_{ab} - K h_{ab})\,,
\end{equation}
meaning that
\begin{equation}
    \mathcal{J}_{ab} = R_{ab} + K \left( \frac{1}{16} [K_{ab} - K h_{ab}] -\frac{9}{16}K_{ab} \right)\,,\qquad \qquad \mathcal{J} = R -\frac{3}{4}K^2\,.
\end{equation}
We now reverse the previous expression and equate both AdS regions inside/outside of the bubble, as they yield the same quantity.
\begin{equation}
    \begin{split}
    \left(\frac{1}{K_+} - \frac{1}{K_-}\right)R_{ab} &= \left[\frac{\mathcal{J}_{ab}}{K}\right]^{+}_{-}-\frac{1}{16}[K_{ab} - K h_{ab}]^+_- + \frac{9}{16}[K_{ab}]^{+}_{-}\\
    &=\left[\frac{\mathcal{J}_{ab}}{K}\right]^{+}_{-} - \frac{1}{16} \ell_5^3 S_{ab} + \frac{9}{16} \ell_5 S_{ab}^\mathrm{TR}
    \end{split}
\end{equation}
Trace reversing the Ricci tensor, one will finally obtain the usual Einstein equation as:
\begin{equation}
    \begin{split}
        \left(\frac{1}{K_+} - \frac{1}{K_-}\right)G_{ab} &= \left[\frac{\mathcal{J}_{ab}}{K}\right]^{+}_{-} -\frac{1}{2}\left[\frac{\mathcal{J}}{K}\right]^{+}_{-}h_{ab}- \frac{3}{8}\left[K\right]^{+}_{-}\,h_{ab} - \frac{1}{16} \ell_5^3 S_{ab} + \frac{9}{16} \ell_5 S_{ab}^\mathrm{TR}\,\\
        &= \left[\frac{\mathcal{J}_{ab}}{K}\right]^{+}_{-} -\frac{1}{2}\left[\frac{\mathcal{J}}{K}\right]^{+}_{-}h_{ab}+ \frac{3}{16}\left[K\right]^{+}_{-}\,h_{ab} + \frac{1}{2} \ell_5^3 S_{ab}\,,
    \end{split}
\end{equation}
where we have made use of eq.~\eqref{eq:trace_reversed} to rewrite the trace-reversed contribution $S^{\text{TR}}_{ab}$. Then using that $K=-4k$ we recover the familiar expression,
\begin{equation}\label{eq:4dEinsteinEq_Exact_small_k}
    \left(\frac{1}{k_+} - \frac{1}{k_-}\right)G_{ab} = \left(\frac{\mathcal{J}_{ab}^+}{k_+} - \frac{\mathcal{J}_{ab}^-}{k_-}\right) -\frac{1}{2}\left(\frac{\mathcal{J}^+}{k_+} - \frac{\mathcal{J}^-}{k_-}\right)h_{ab}+ 3(k_+ - k_-)h_{ab} - 2 \ell_5^3 S_{ab}\,.
\end{equation}
Note that this did not require any approximations. These are precisely the 4d Einstein equations from before.

\bibliography{biblio}
\bibliographystyle{utphysmodb}

\end{document}